\documentclass[11pt]{article}
\usepackage{graphicx}
\usepackage{epstopdf}
\usepackage{amsmath}
\usepackage{amssymb}
\usepackage{verbatim}
\usepackage{algorithmic}
\usepackage{subfig}
\usepackage{bm}

\begin{document}

\title{Magnetic stripe domain pinning and reduction of in plane magnet order due to periodic defects in thin magnetic films}

\author{R. L.~Stamps$^{1,2}$ and M. C.~Ambrose $^{1}$}
\maketitle
\noindent $^{1}$ School of Physics, The University of Western Australia, 35 Stirling Hwy, Crawley 6009, Australia
\\ $^{2}$ SUPA School of Physics and Astronomy, University of Glasgow, Glasgow G12 8QQ, United Kingdom
\date{\today}
\section*{Abstract}
pacs: 75.30.Kz,64.70.dj,07.05.Tp,75.30.Cr
\\
In thin magnetic films with strong perpendicular anisotropy and strong demagnetizing field two ordered phases are possible. At low temperatures, perpendicularly oriented magnetic domains form a striped pattern. As temperature is increased the system can undergo a spin reorientation transition into a state with in-plane magnetization. Here we present Monte Carlo simulations of such a magnetic film containing a periodic array of non-magnetic defects. We find that the presence of defects stabilizes parallel orientation of stripes against thermal fluctuations at low temperatures. Above the spin reorientation temperature we find that defects favor perpendicular spin alignment and disrupt long range ordering of spin components parallel to the sample. This increases cone angle and reduces in plane correlations, leading to a reduction in the spontaneous magnetization.

\section{Introduction}
Quasi two dimensional ultra thin magnetic films engender a large area of theoretical and technical interest, due in part to the large variety of magnetic properties that can be produced \cite{Jensen2006129,PhysRevLett.56.2728} and their applications in data storage \cite{5389136,4900605}.
For a sufficiently high ratio of dipole to exchange coupling strengths, the ground state of thin magnetic films can consist of magnetic stripe domains \cite{rossol:5263, PhysRevLett.84.2247,PhysRevB.68.212404}. For films with a strong perpendicular anisotropy a second phase transition is possible, in which spins reorient resulting in a non zero magnetization parallel to the sample plane \cite{Politi1995647,PhysRevLett.69.3385, PhysRevLett.64.3179}.\\
There are a number of lithographic techniques that can be used to create nanometer scale magnetic structures \cite{1367-2630-11-12-125002,4717577,Li20118307,smyth:4237,doi:10.1021/nl300622p,4964274,so78097,doi:10.1021/ar980081s,Martı́n2003449}. When compared with isotropic films, periodic magnetic nano structures have been shown to significantly alter macroscopic properties such as anisotropy \cite{castano:2872,vavassori:7992}, magneto-resistance \cite{castano:2872}, cohesive field \cite{moore:8746,Kronmüller1981159} and spin reorientation temperature \cite{PhysRevB.86.094431,bergeard:103915}.
 \\
On the micro scale, magnetic stripe domains can appear with long range orientaional order \cite{PhysRevB.68.212404, PhysRevLett.84.2247,PhysRevLett.93.117205} or forming complex patterns  \cite{PhysRevB.55.2752,PhysRevLett.84.2247}. Nano scale patterning has been to used to create pinning sites for domain walls \cite{cowburn:2309,perez-junquera:033902,0022-3727-40-10-006,metaxas:132504, konings:033904,PhysRevB.86.094431}. When the period of pinning cites is comparable to the natural stripe width, long range orientational order can be stabilized \cite{konings:054306,PhysRevB.86.094431} .
\\
Theoretically these quasi two dimensional systems have been studied with a variety of methods. For two dimensional isotropic systems the problem of melting is reasonably well understood \cite{RevModPhys.60.161}, in particular the spin reorientation transition and stripe melting have been studied analytically \cite{PhysRevB.42.849,PhysRevLett.65.2599, PhysRevB.48.10335,PhysRevB.51.1023} and with computer simulation \cite{PhysRevB.77.134417, PhysRevB.77.174415,2paper}. Theoretically the problem of melting in two dimensional systems has been considered for the case of particles with a periodic potential \cite{PhysRevB.19.2457}. The pinning of domain walls has been explored for both random \cite{0022-3727-42-12-125001} and periodic defects\cite{PhysRevB.67.115412}. Recently micro-magnetic computer simulations have explored the contribution of periodic defects and edge effects to magnetic reversal and hysteresis \cite{wiele:053915}. Here we perform Monte Carlo simulations on striped magnetic system in order understand the effect of periodic non magnetic defects on the thermally driven spin reorientation and stripe melting transitions.
\section{Method}
The system is modeled as a two dimensional square array of Heisenberg spins $\bm{s}_{\bm{i}} \in \mathcal{S}^2$, with lattice spacing $\alpha$.
\begin{equation}
\begin{split}
 H= &\frac{J}{2}\sum_{\langle \bm{i},\bm{j} \rangle} \bm{s}_{\bm{i}} \cdot \bm{s}_{\bm{j}} +K \sum_{\bm{i}} (s_{\bm{i}}^z)^2  \\
 +&\frac{C_D}{2} \sum_{\bm{i},\bm{j}} \frac{1}{r_{\bm{\bm{ij}}}^3} (\bm{s}_{\bm{i}} \cdot \bm{s}_{\bm{j}} -3 \bm{s}_{\bm{i}} \cdot \hat{\bm{r}}_{\bm{\bm{ij}}}\bm{s}_{\bm{j}} \cdot 
\hat{\bm{r}}_{\bm{\bm{ij}}})
\end{split}
\end{equation}
where $\bm{i}$ and $\bm{j}$ represent two dimensional indexes, $\bm{i} = (i_x, i_y)$, $s_{\bm{i}}^z = \bm{s}_{\bm{i}} \cdot \hat{\bm{z}} $  and $\langle \dots \rangle$ indicates the sum extends only over nearest neighbors. $J$, $K$ and $C_D$ represent the strength of the exchange coupling, perpendicular anisotropy and dipole coupling respectively. In order to introduce non-magnetic defects, some lattice sites are left empty. These defects are arranged as a regular square array with spacing $w_d$.
The system is evaluated with Metropolis algorithm Monte Carlo. In order to approximate an infinite system, periodic boundary conditions are introduced. After the change of co-ordinates $\bm{r}_{\bm{nm}} = \bm{G}+\bm{\rho}_{\bm{n'm'}}$ where $\bm{\rho}_{\bm{n'm'}} = (\rho_x,\rho_y)$ and $\rho_x,\rho_y \in [0,L]$, dipole coupling is calculated over a series of replicas of the original system \cite{Harris, Ewald} the dipole interaction and Monte Carlo steps are parallelized on a GPU using the stream processing method described in our previous paper \cite{2paper}. Non magnetic sites $\bm{s}_{\bm{i}} = \bm{0}$ are not updated.
\section{Results}
\label{Res}
\begin{figure}[!htb]
  \centering
  \includegraphics[width=7cm]{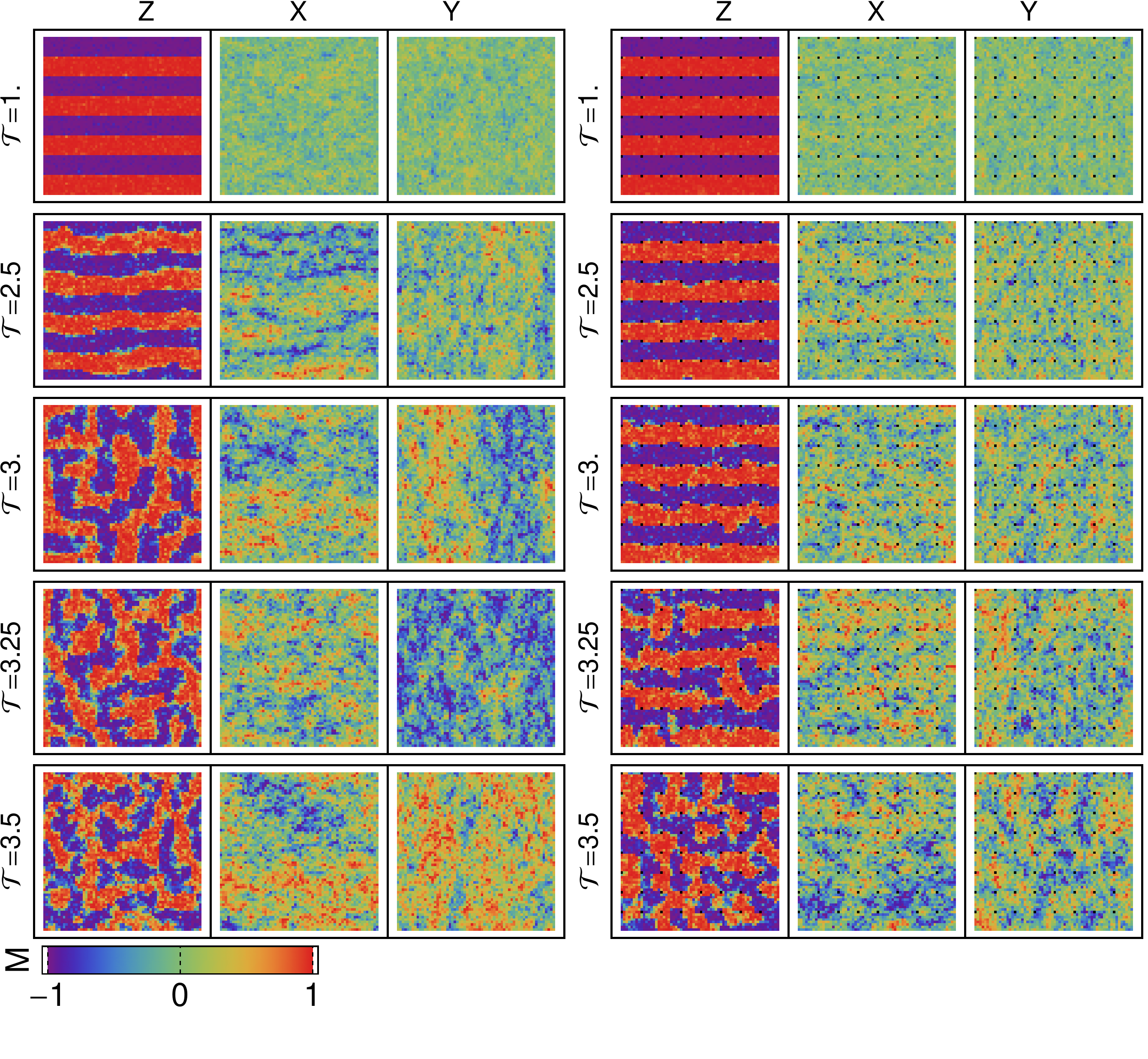}
  \caption[Stripe Pinning]%
  {Example of spin configurations near the loss of orientational order. Spins vales are indicated according the color scale shown below, defects are colored black. Columns from left to right: $s_{\bm{i}}^z$, $s_{\bm{i}}^x$ and $s_{\bm{i}}^y$ followed by the same states in the presence of defects. Rows from top to bottom: $\mathcal{T} = 1$, $\mathcal{T} = 2.5$, $\mathcal{T} = 3$, $\mathcal{T} = 3.25$ and $\mathcal{T} = 3.5$ }
\label{OTE}
\end{figure}

\begin{figure}[!htb]
  \centering
  \includegraphics[width=7cm]{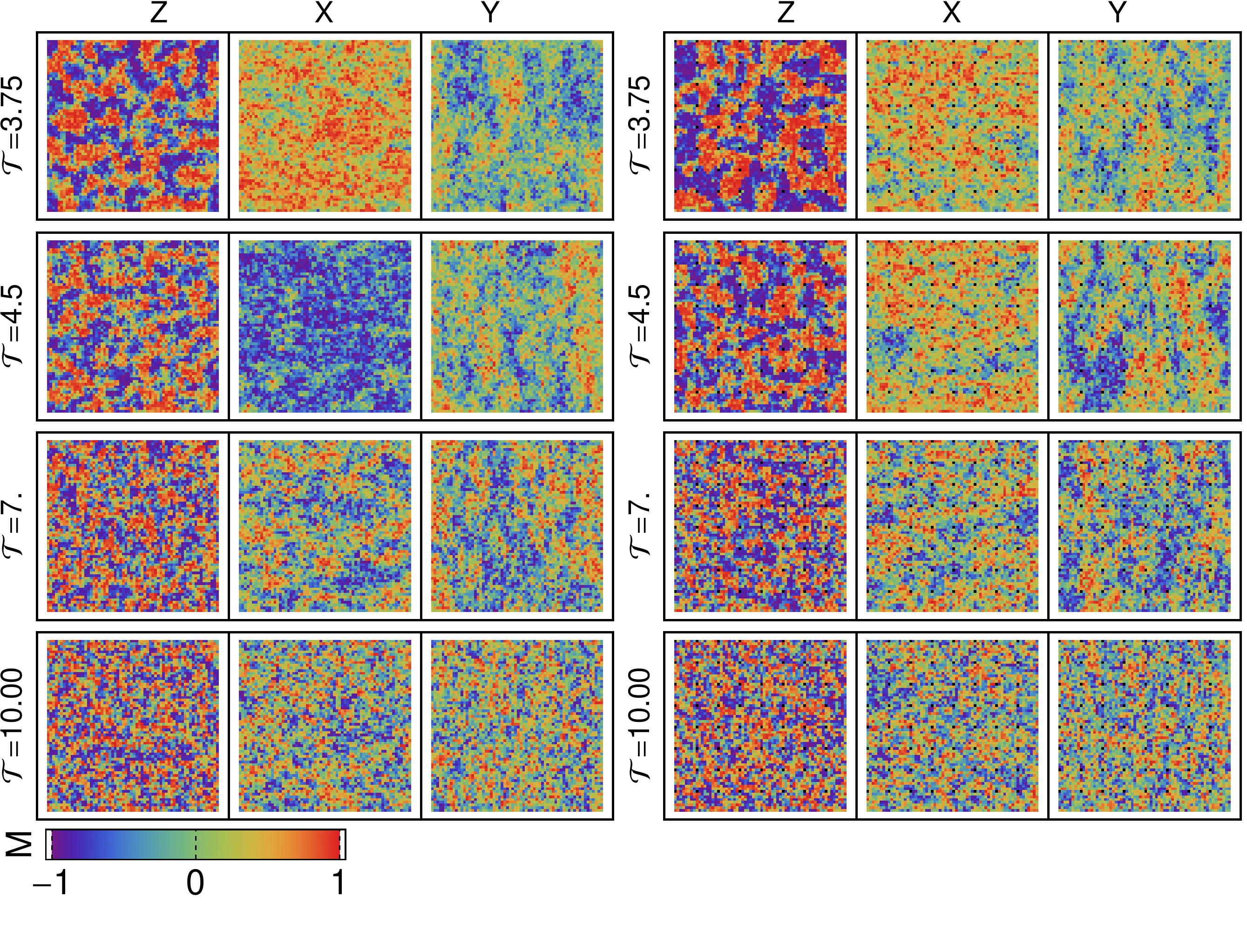}
  \caption[Melting with Defects]%
  {Example of spin configurations at high temperature. Colors are used to indicate spins values and the loation of defects as in Fig. \ref{OTE}. Columns from left to right: $s_{\bm{i}}^z$, $s_{\bm{i}}^x$ and $s_{\bm{i}}^y$ followed by the same states in the presence of defects.Rows from top to bottom: $\mathcal{T} = 3.75$ , $\mathcal{T} = 4.5$ ,$\mathcal{T} = 7$ and $\mathcal{T} = 10$ }
\label{OTHi}
\end{figure}

In order to create a periodic array of defects we select a system size of $L=64 \alpha$ and defect spacing $w_d = 8 \alpha$. The ratio $\mathcal{K}=K/C_D$ was set as $\mathcal{K}=15$ ensuring that the ground state was not canted ($\bm{s}_{\bm{i}} \cdot \hat{\bm{z}} =\pm 1$). The ratio of exchange to dipole coupling is selected to be $\mathcal{J}=J/C_D= 8.9$ giving a stripe width of $w_s = 8 \alpha$. At $T=0$, when the system is ordered, we find that for the choice of parameters above, the lowest energy occurs when domain boundaries pass through magnetic defects (this minimizes the energy by replacing a high energy spin with a defect). The system is initiated in the ground state and Monte Carlo ensembles are generated disregarding the initial $10^5$ Monte Carlo steps to allow the system to equilibrate. A further $5 \times 10^4$ steps are taken with states recorded every $50$ steps. Previously we determined that $50$ steps allowed sufficient independence between ensemble configurations. In order to examine the effects of the defects results are included from an identical simulation performed on a perfect lattice that we shall refer to as the isotropic case.
\\
When describing results we will refer to the normalized temperature $\mathcal{T}=k_B T C_D^{-1}$. In Fig. \ref{OTE} sample states are shown for low temperatures near to where orientational order is destroyed. In the absence of defects, as $\mathcal{T}$ is increased, the striped system initially undergoes roughening at the stripe boundaries. The roughened domain boundaries are associated with localized canting of the spins away from perpendicular alignment. As temperature is further increased the system undergoes bridging between stripes that leads to the destruction of long range orientational order. With the inclusion of defects the same general trends occur: stripe roughening followed by bridging and eventual destruction of long range order. However, the presence of defects stabilizes the striped order at higher temperatures. In addition, differences in morphology are observed. In the absence of defects the stripes display long wavelength undulations. In the presence of defects walls are pinned. Instead of long wavelength bending, fluctuations exist as roughening of the sections of wall between defects. Also, in contrast to the isotropic case, we observe that this initial roughening of stripes is not associated with the appearance of canted spins. 
\\
In Fig. \ref{OTHi} the behavior of the two systems is shown at temperatures above the loss of orientational order. In both cases the system forms regions with spins canted towards in-plane alignment and the existence of long range order in the in-plane components. As temperature is increased the systems become increasing granular before reaching the paramagnetic limit.
\subsection{Orientational Order Parameter}
\begin{figure}[!htb]
  \centering
  \includegraphics[width=6cm]{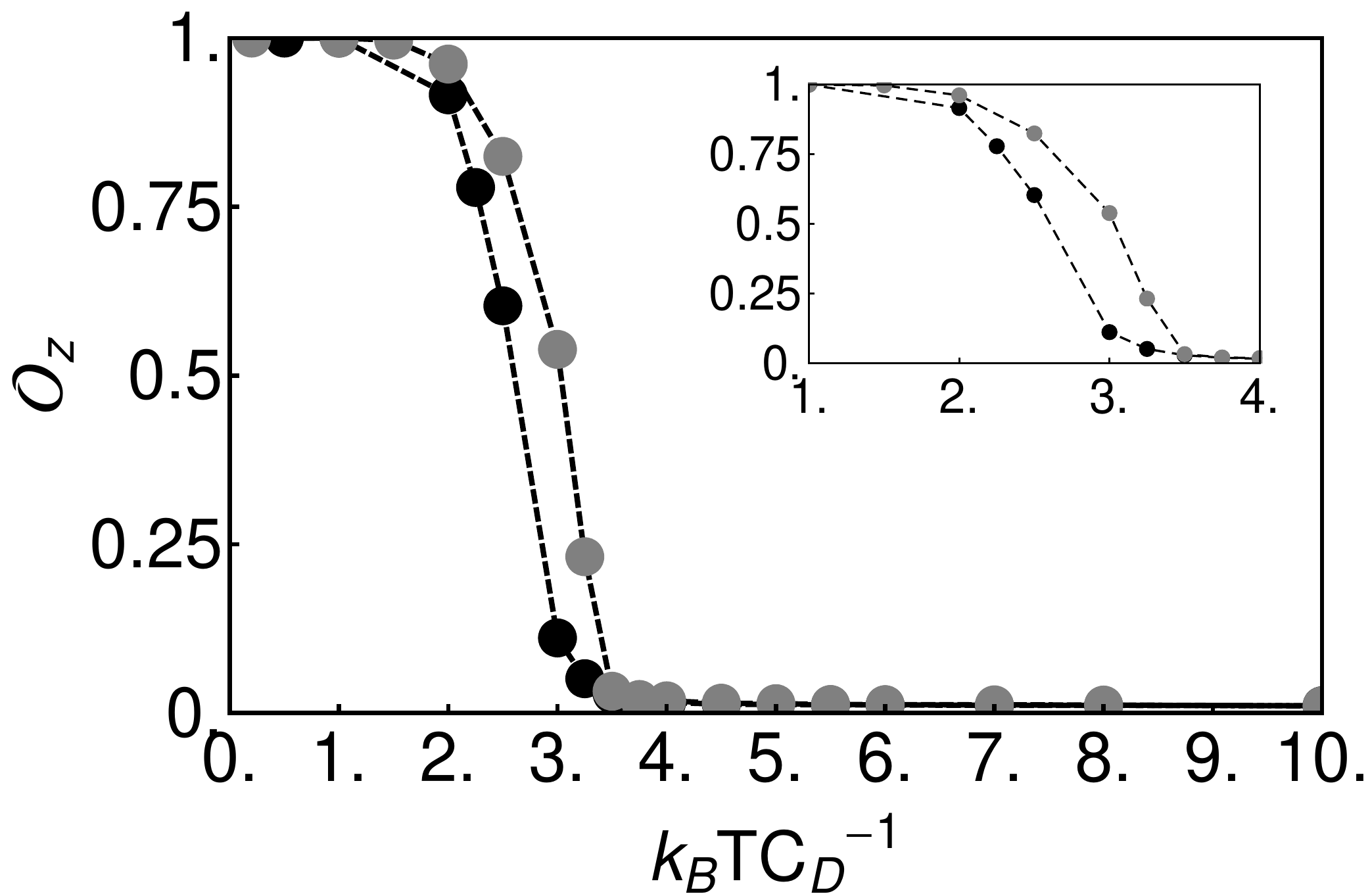}
  \caption[Orientational Order Parameter]%
  {Orientational Order Parameter as a function of $\mathcal{T}$, black circles represent the uniform system, while gray dots represent the system in the presence of defects. The transition region is replotted with a finer $\mathcal{T}$ scale in the insert.}
\label{OT}
\end{figure}

In order to analyze the loss of orientational order we locate vertical and horizontal perpendicular domain walls by using $n_h^z$ and $n_v^z$ \cite{PhysRevB.77.174415, PhysRevLett.75.950,2paper}
\begin{equation}
\label{HVO}
\begin{split}
n_{h}^{z} &= \frac{1}{2N} \sum_{\bm{i,j} \: \text{v.n.n}}  1-\text{sgn}(\bm{s}_{\bm{i}} \cdot \hat{\bm{z}} \; \bm{s}_{\bm{j}} \cdot \hat{\bm{z}}) \\
n_{v}^{z} &= \frac{1}{2N} \sum_{\bm{i,j} \: \text{h.n.n}}  1-\text{sgn}(\bm{s}_{\bm{i}} \cdot \hat{\bm{z}} \; \bm{s}_{\bm{j}} \cdot \hat{\bm{z}})
\end{split}
\end{equation}
where v.n.n and h.n.n indicate that the sums should be taken over all pairs of spins which are nearest neighbors in the horizontal and vertical directions respectively. The orientational order is given by
\begin{equation}
\mathcal{O}_z = \left \langle |n_h^z-n_v^z|/(n_h^z+n_v^z) \right \rangle
\end{equation}
With the inclusion of defects the sums in Eq.  \ref{HVO} are restricted to run over all pairs that are not defects.
\\
In Fig. \ref{OT} $\mathcal{O}_z$ is plotted as a function of the normalized temperature $\mathcal{T}$. 
At low $\mathcal{T}$ both systems display a striped array with smooth boundaries corresponding to $\mathcal{O}_z = 1$. We observe that, while the transition profile is similar, the presence of defects increases the transition temperature.
\subsection{In Plane Magnetic Order}
\begin{figure}[htb]
  \centering
  \includegraphics[width=6cm]{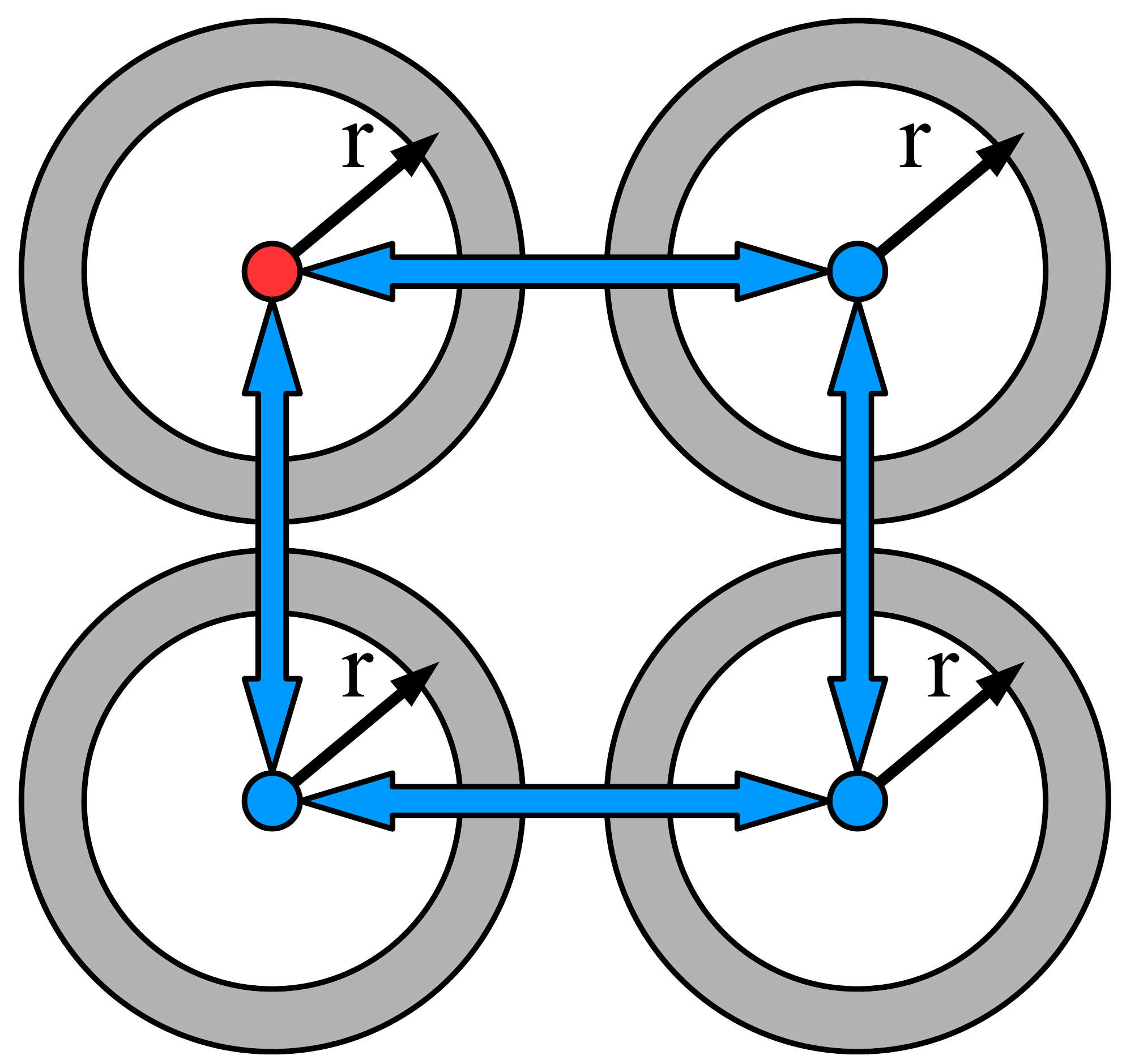}
  \caption[Parallel Magnetization as a function of $\mathcal{T}$]%
  {$G_{\bm{i}}(r)$ calculates the average correlation between the spin at site $\bm{i}$ (red circle) with all spins within a fixed radius (upper left gray circle). Spins separated by integer combinations of the vectors $w_d \hat{\bm{x}}$ $ w_d \hat{\bm{y}}$ will have equivalent proximity to their nearest defect. $\mathcal{G}_{\bm{i}}(r)$ averages $G_{\bm{i}}(r)$ over these equivalent sites.}
\label{exp}
\end{figure}

\begin{figure}[htb]
  \centering
  \includegraphics[width=6cm]{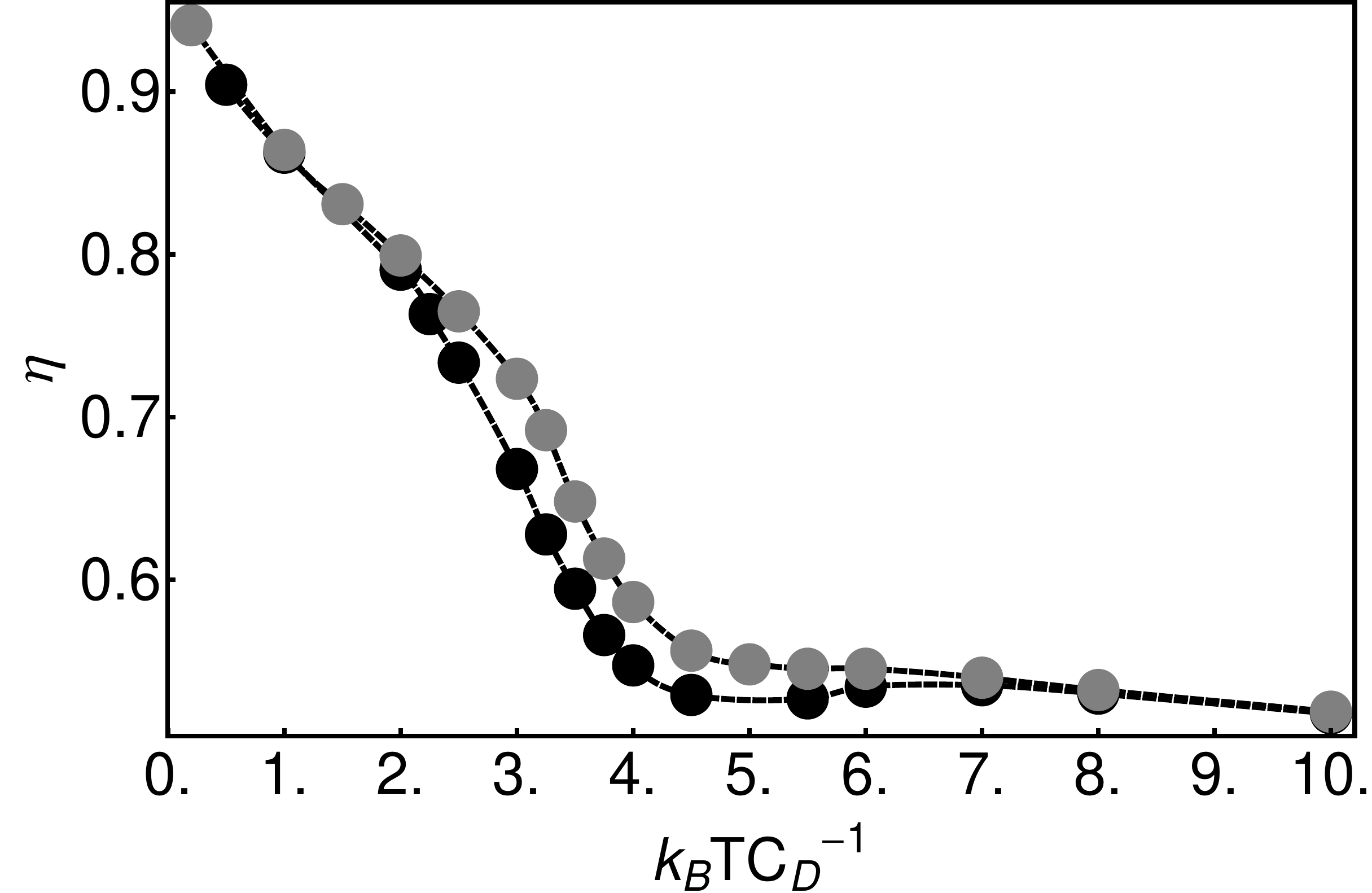}
  \caption[$\eta$ in the presence of defects]%
  {$\eta$ as a function of $\mathcal{T}$, black circles represent the uniform system, while gray dots represent the system in the presence of defects.}
\label{eta}
\end{figure}

\begin{figure}[!h]
  \centering
  \includegraphics[width=8cm]{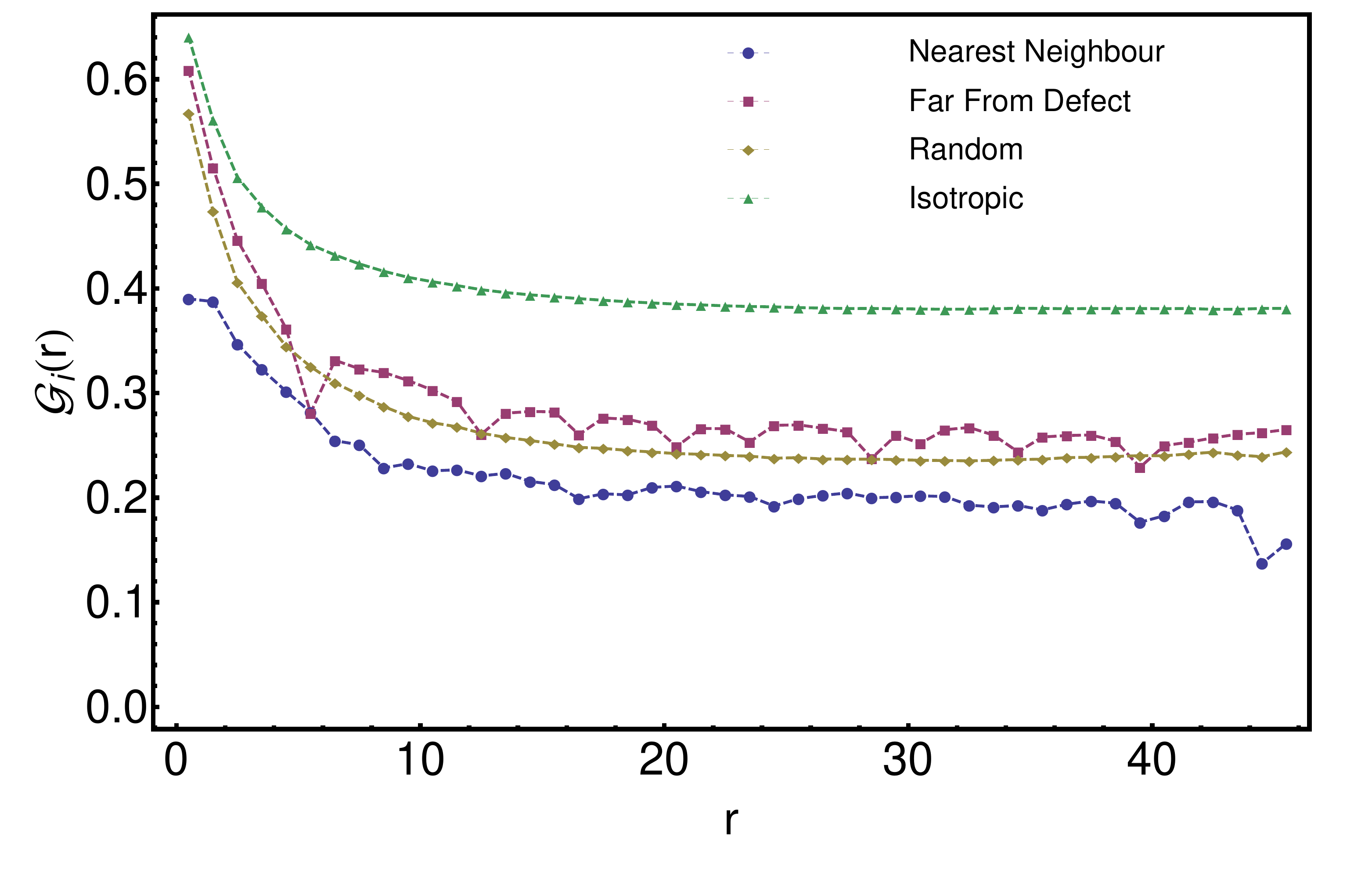}
  \caption[Correlation function in the presence of defects]%
  {$\mathcal{G}_{\bm{i}}$ as a function of $r$ (given here in units of $\alpha$) at $\mathcal{T} = 4.5$ for $\bm{i}$; as a nearest neighbor to ordered defects (blue circles), $\bm{i}$ at a maximum distance from ordered defects (red squares), in the presence of random defects (yellow diamonds) and for a defect free case (green triangles). }
\label{Corlen}
\end{figure}

\begin{figure}[htb]
  \centering
  \includegraphics[width=6cm]{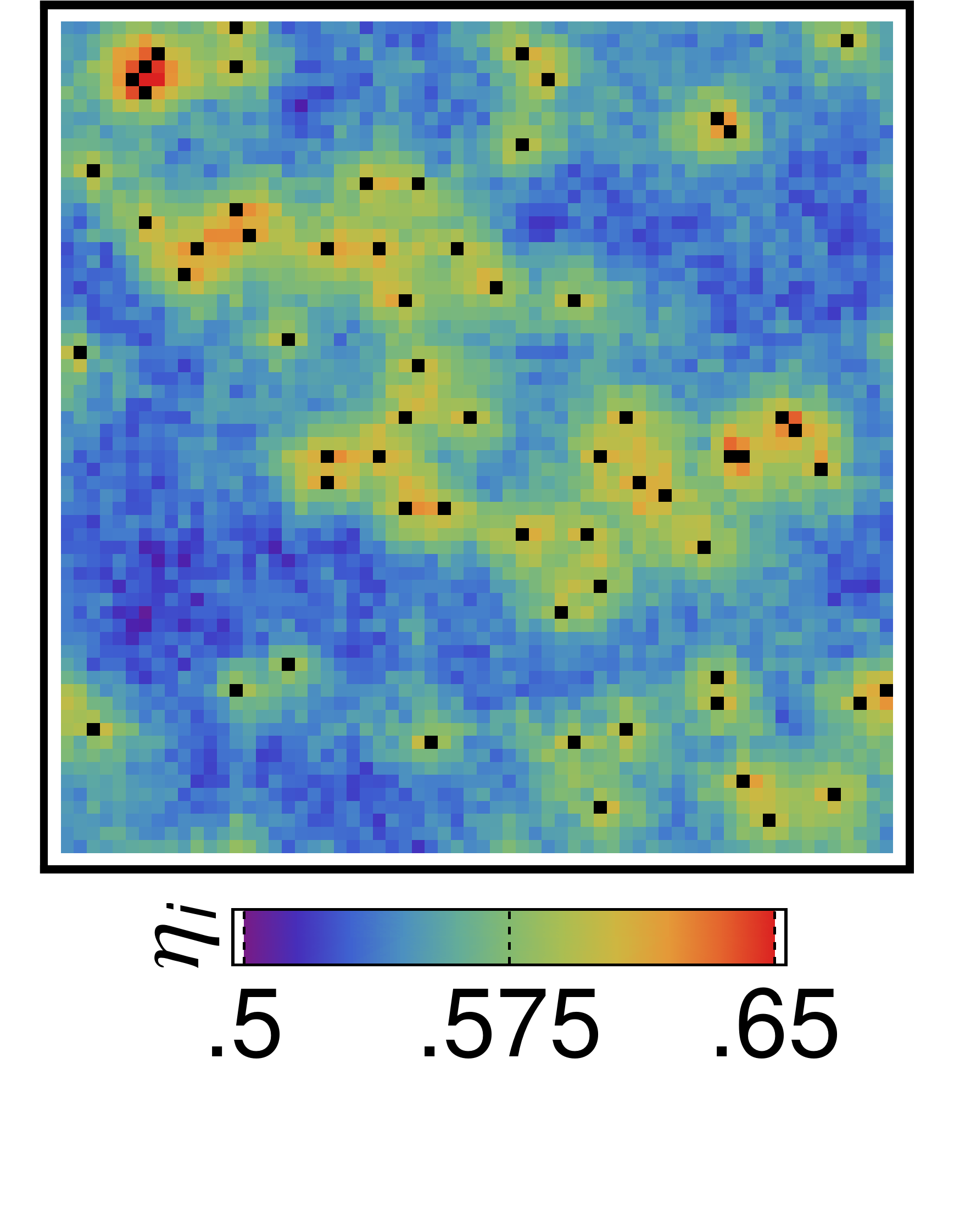}
  \caption[Spatial dependence of $\eta$ in the presence of defects]%
  {$\langle \eta_{\bm{i}} \rangle$ in the presence of randomly spaced defects at $\mathcal{T} = 4.5$, black squares indicate defects and the value of $\eta_{\bm{i}}$ is indicated on the scale below.}
\label{eta_space_rand}
\end{figure}

\begin{figure}[htb]
  \centering
  \includegraphics[width=6cm]{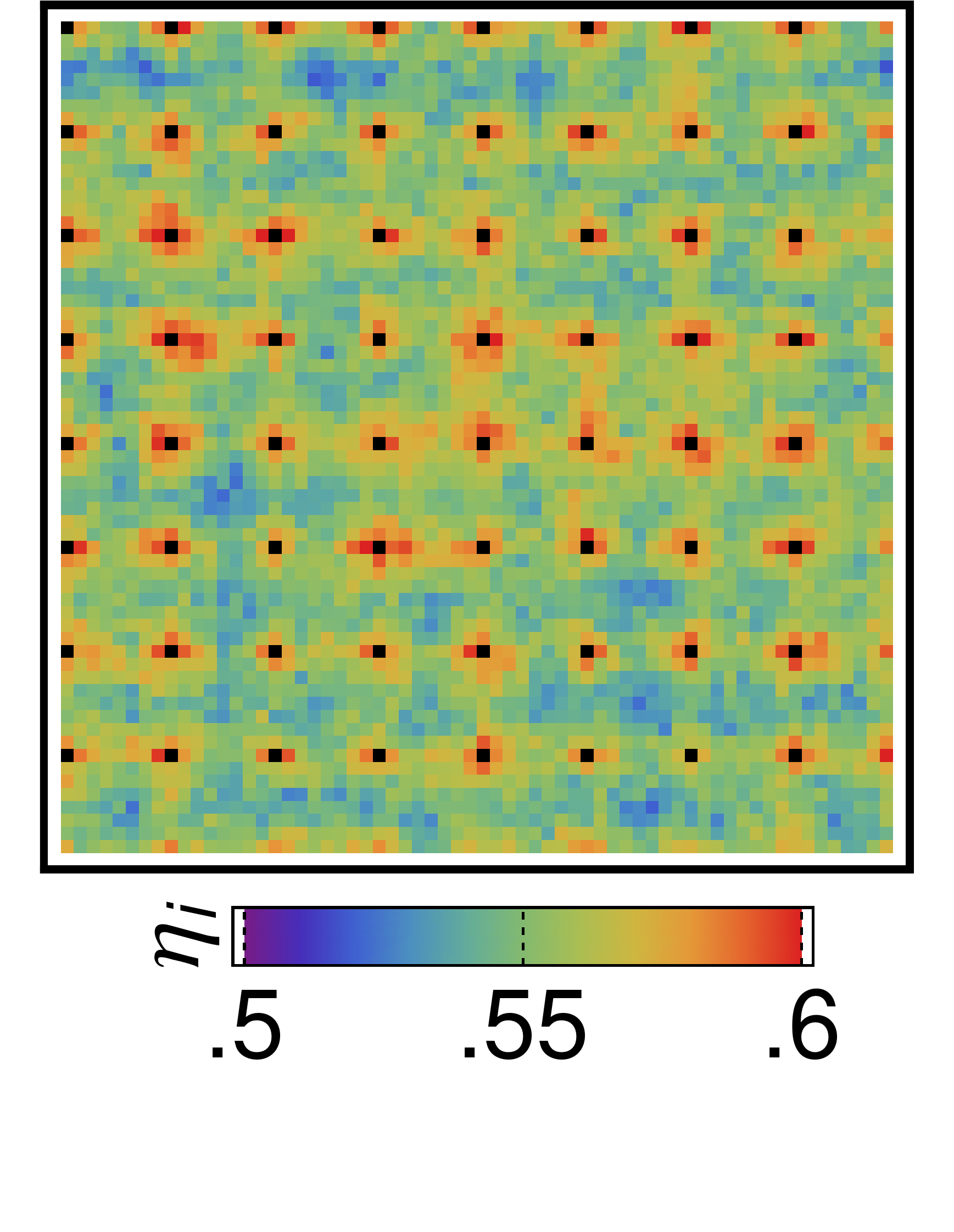}
  \caption[Spatial dependence of $\eta$ in the presence of defects]%
  {$\langle \eta_{\bm{i}} \rangle$ in the presence of regularly spaced defects at $\mathcal{T} = 4.5$, black squares indicate defects and the value of $\eta_{\bm{i}}$ is indicated on the scale below.}
\label{eta_space}
\end{figure}

\begin{figure}[!h]
  \centering
  \includegraphics[width=8cm]{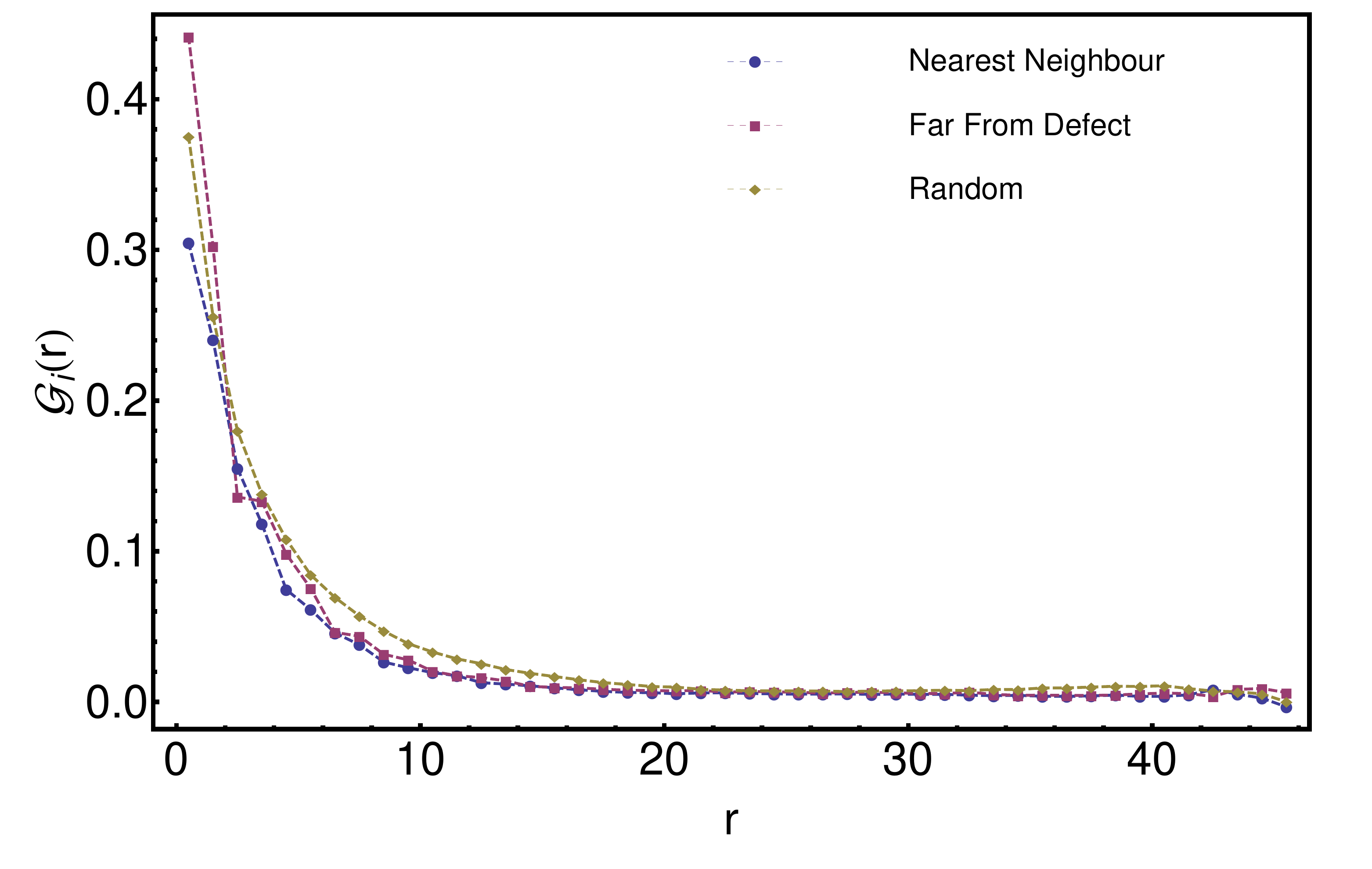}
  \caption[Correlation function in the presence of defects]%
  {$\mathcal{G}_{\bm{i}}$ as a function of $r$ (given here in units of $\alpha$) at $\mathcal{T} = 4.5$ in the presence of a high density of defects for $\bm{i}$; as a nearest neighbor to ordered defects (blue circles), $\bm{i}$ at a maximum distance from ordered defects (red squares) and in the presence of random defects (yellow diamonds). }
\label{CorlenHD}
\end{figure}

\begin{figure}[htb]
  \centering
  \includegraphics[width=6cm]{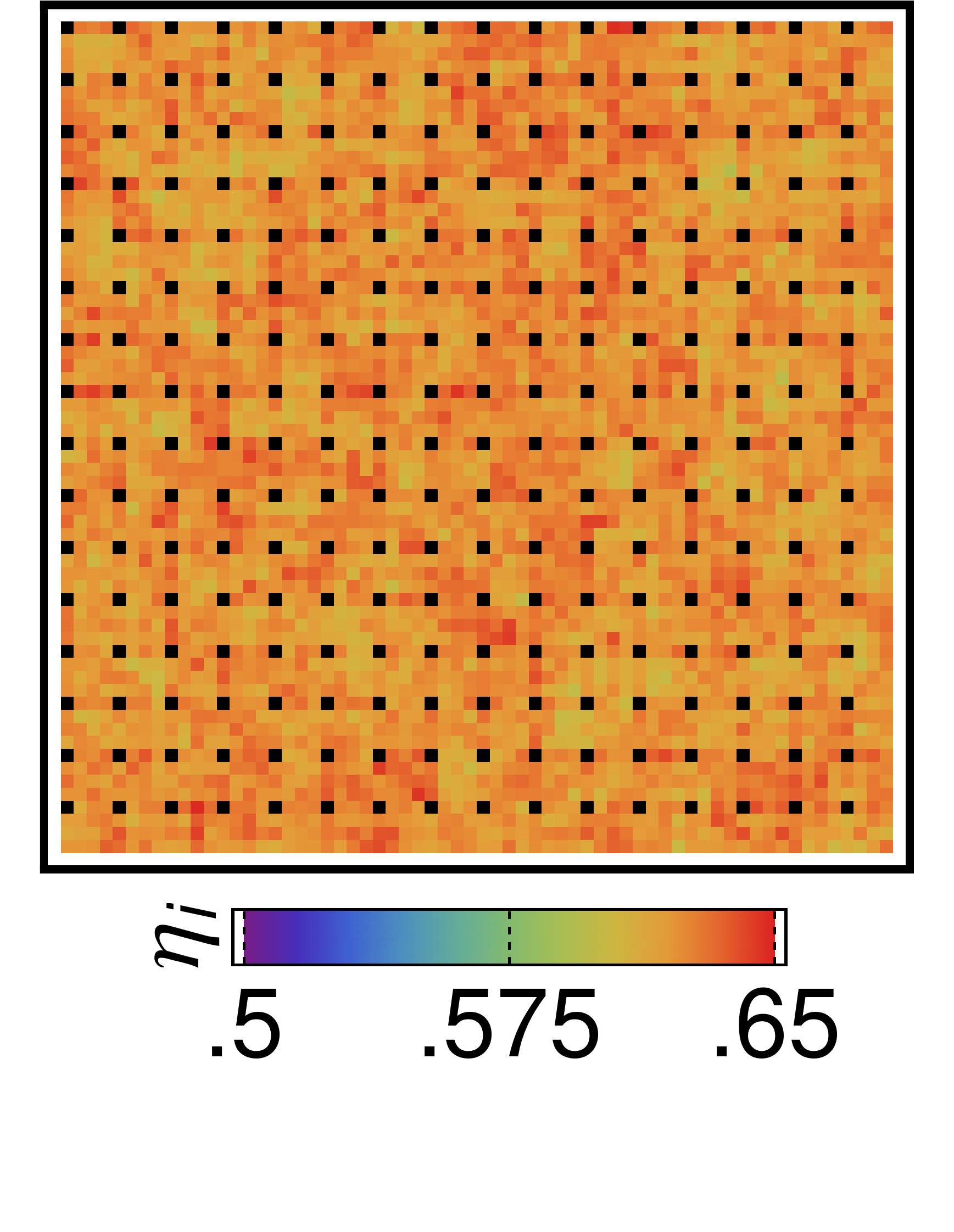}
  \caption[Spatial dependence of $\eta$ in the presence of defects]%
  {$\langle \eta_{\bm{i}} \rangle$ in the presence of closely spaced defects at $\mathcal{T} = 4.5$, black squares indicate defects and the value of $\eta_{\bm{i}}$ is indicated on the scale below.}
\label{eta_spaceHD}
\end{figure}

\begin{figure}[htb]
  \centering
  \includegraphics[width=6cm]{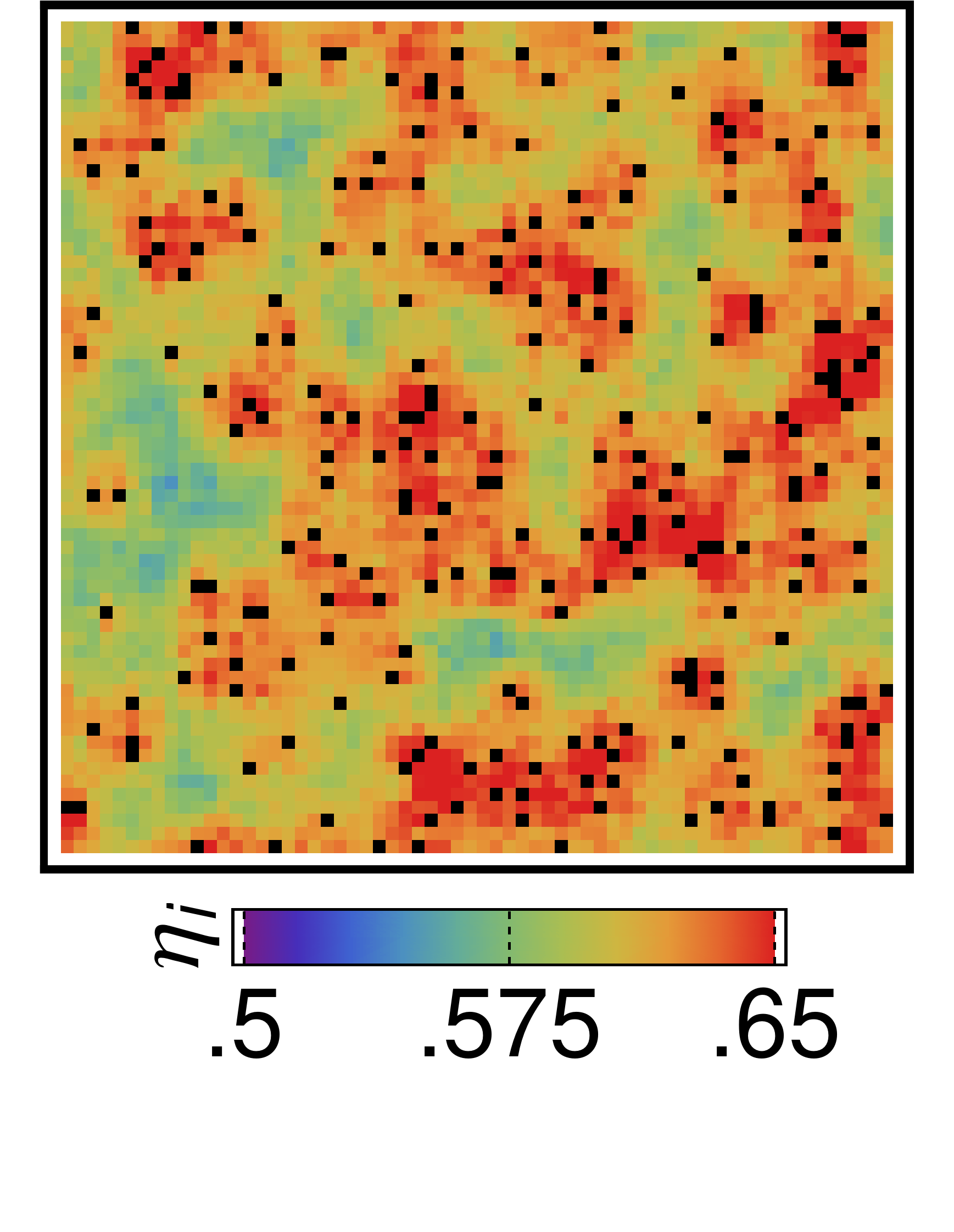}
  \caption[Spatial dependence of $\eta$ in the presence of defects]%
  {$\langle \eta_{\bm{i}} \rangle$ in the presence of regularly spaced defects at $\mathcal{T} = 4.5$, black squares indicate defects and the value of $\eta_{\bm{i}}$ is indicated on the scale below.}
\label{eta_space_randHD}
\end{figure}

At high temperature when spins are no longer entirely perpendicular the system can display net magnetization parallel to the system plane. Letting $M_x = 1/N \sum_{\bm{i}} s_{\bm{i}}^x$ and $M_y = 1/N \sum_{\bm{i}} s_{\bm{i}}^y$ (with $N=L^2$) , the in plane magnetization is 
\begin{equation}
M_{\parallel} = \langle(M_x^2+M_y^2)^{\frac{1}{2}} \rangle .
\end{equation}
In plane magnetic order can occur only when spins are canted away from the perpendicular alignment, in order to measure the degree of canting we use the cone angle
\begin{equation}
\begin{split}
\eta &= \left  \langle \frac{1}{N} \sum_{\bm{i}} \eta_{\bm{i}} \right \rangle \\
\text{with}\\
\eta_{\bm{i}} &=\sqrt{ (2/\pi)^{2} \langle(\theta_{\bm{i}}-\pi/2)^2 \rangle}
\end{split}
\end{equation}
where $\theta_{\bm{i}}$ is the zenith angle of the spin at site $\bm{i}$. When calculating $\eta$ and $M_{\parallel}$ in the presence of defects $N$ is replaced with $N' = N (w_d^2-1)/w_d^2$ to account for the fact that the defects don't contribute to the averages. In addition to these two single site order parameters, we calculate correlation functions between different spins. Taking $\theta_{\bm{i}}$ and $\phi_{\bm{i}}$ as the zenith and azimuthal angles of $\bm{s}_{\bm{i}}$ respectively, $\bm{s}_{\bm{i}}.\bm{s}_{\bm{j}} = \cos(\theta_{\bm{i}})\cos(\theta_{\bm{j}}) + \sin(\theta_{\bm{i}})\sin(\theta_{\bm{j}})\cos(\phi_{\bm{i}}-\phi_{\bm{j}})$. Since we are interested in plane ordering we calculate
\begin{equation}
G_{\bm{ij}}= \begin{cases}
\langle \cos(\phi_{\bm{i}}-\phi_{\bm{j}}) \rangle & \text{for $\bm{s}_{\bm{i}} \ne 0 $ and $\bm{s}_{\bm{j}} \ne 0 $ } \\
  0 & \text{otherwise}
\end{cases}
\end{equation}
In order to calculate the correlation as a function of distance we define
\begin{equation}
\begin{split}
G_{\bm{i}}(r) = \frac{1}{N_r} \sum_{\bm{j}} \Pi((r_{\bm{ij}} -r)/\alpha ) G_{\bm{ij}}
\end{split}
\end{equation}
\begin{figure}[htb]
  \centering
  \includegraphics[width=4cm]{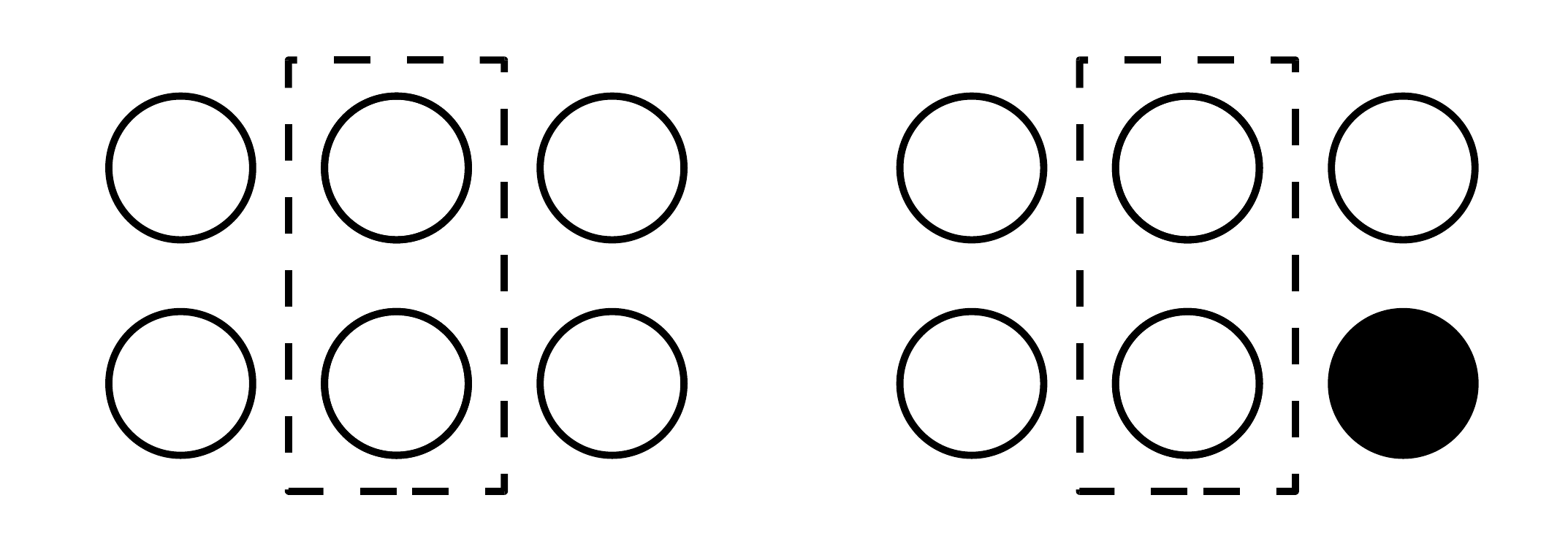}
  \caption[Defect proximity alters correlation strength]%
  {The correlation between the circled sites on the left will not be equal to the correlation between the circled sites on the right due to the presence of a defect (indicated here by a black circle).}
\label{CorCom}
\end{figure}
Here $\Pi$ is the Heaviside Pi function ($\Pi(x)= \Theta(x+1/2)\Theta(x-1/2)$) and $N_r$ is the number of spins contained in the average $G_{\bm{i}}(r)$; $N_r = \sum_{\bm{i}} \Pi((r_{\bm{ij}} -r)/\alpha )$.
 $G_{\bm{i}}(r)$ is the average correlation of the spin at site $\bm{i}$ with spins at a radius $r$ from $\bm{i}$. In a spatially isotropic state one expects that $G_{\bm{i}}(r)$ should depend only on the separation between spins. Here the inclusion of periodic defects breaks the isotropy. The correlation between two spins separated by distance $r$ will depend on the proximity of  the spins to a defect. We define the following average; letting $n = (L/w_d)$
\begin{equation}
\begin{split}
\mathcal{G}_{\bm{i}}(r) &= \frac{1}{n^2} \sum_{\bm{i}' = 1} G_{\bm{i}'}(r) =   \frac{1}{n^2 N_r} \sum_{\bm{i}' = 1} \sum_{\bm{j}} \Pi((r_{\bm{i}'\bm{j}} -r)/ \alpha ) G_{\bm{i}'\bm{j}} \\
\text{with}\\
\bm{i}' &= \bm{i} + w_d a  \hat{\bm{x}}+ w_d b  \hat{\bm{y}} \text{ for } a,b \in [1,n].
\end{split}
\end{equation}
The meaning of this correlation function is elucidated in Fig. \ref{exp}. Here $G_{\bm{i}}(r)$ calculates the correlation between a fixed spin at site $\bm{i}$  and spins at some fixed distance $r$. Since the system is not spatially isotropic we expect that  $G_{\bm{i}}(r)$ will depend on $\bm{i}$. $\mathcal{G}_{\bm{i}}(r) $ averages the $G_{\bm{i}}(r)$ over all sites with equivalent proximity to their closest defect. In the absence of the symmetry breaking defects and in a uniform phase $\mathcal{G}_{\bm{i}}(r) $ is not dependent on $\bm{i}$.
\begin{figure}[htb]
  \centering
  \includegraphics[width=6cm]{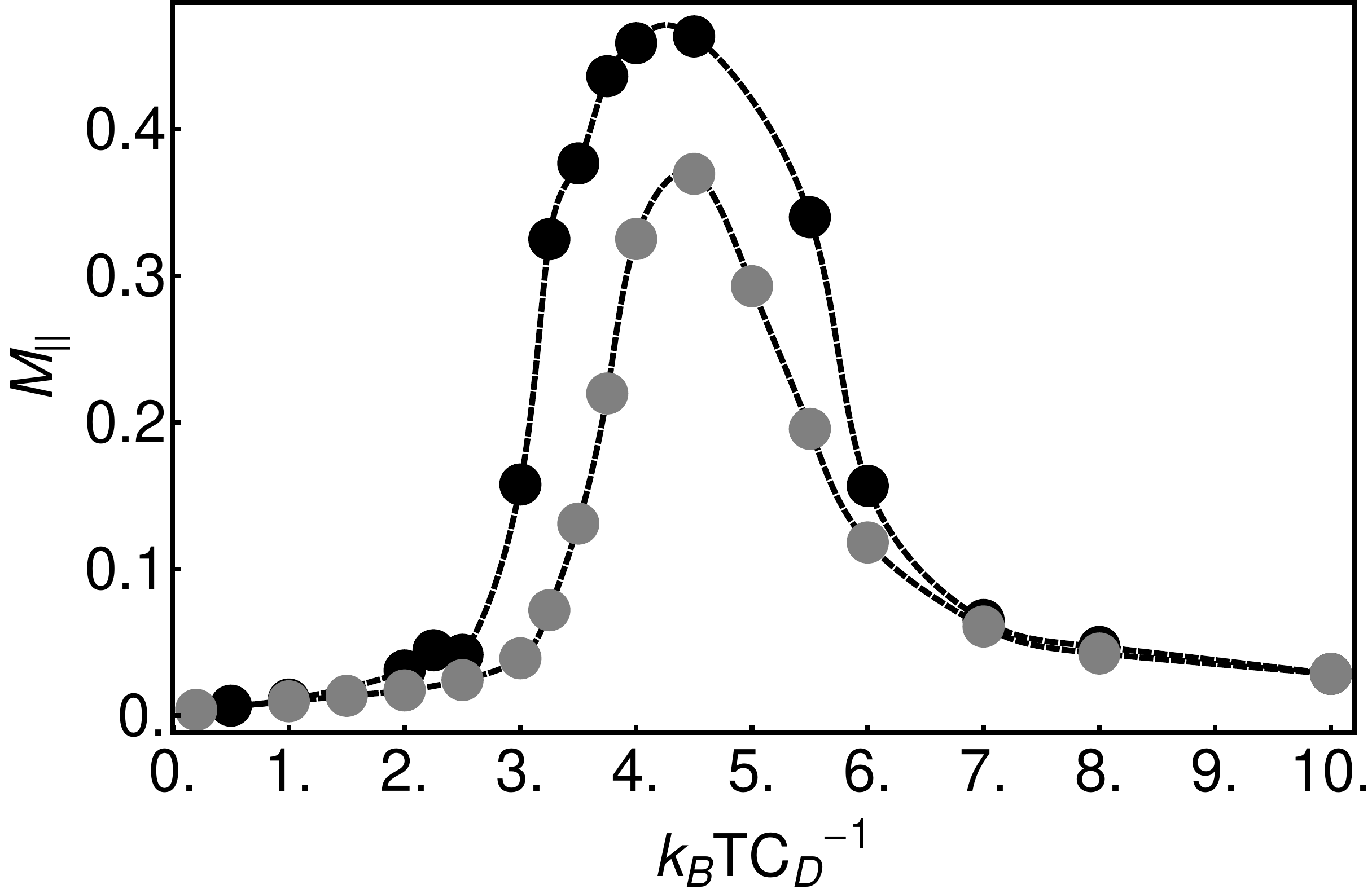}
  \caption[Parallel Magnetization as a function of $\mathcal{T}$]%
  {Parallel magnetization as a function of $\mathcal{T}$, black circles represent the uniform system, while gray dots represent the system in the presence of defects.}
\label{MP}
\end{figure}
\\
In Fig. \ref{MP} the parallel magnetization is shown as a function of temperature. Here we see that in the presence of defects the magnetic ordering is suppressed, and that the peak magnetization is reduced by around $20\%$. In Fig. \ref{eta} we see that the degree of spin canting is reduced for $2<\mathcal{T}<7$, however this reduced spin canting is not sufficient to account for the reduction in peak magnetization. In Fig. \ref{Corlen}, $\mathcal{G}_{\bm{i}}(r)$ is plotted for $\mathcal{T}=4.5$, corresponding to the peak in-plane magnetization. In the isotropic system $\mathcal{G}_{\bm{i}}$ has slow monotonic decay with increasing distance between spins. For the non isotropic system $\mathcal{G}_{\bm{i}}$ is calculated for two choices of $\bm{i}$. The first choice is  $\bm{i}$ as a nearest neighbor to a defect, in this case the correlation is strongly reduced for all $r$. The other choice is $\bm{i}$ at maximum distance from a defect, in this case the correlation is comparable to the isotropic case for small distances. However the correlation strength decreases rapidly as $r$ approaches $r=6\alpha$ (the location of the closest defects).
\\
In addition to the reduction in correlation strength we observe a periodic structure in both the defect cases due to the periodic defect lattice. This effect is particularly strong for the case when the $\mathcal{G}_{\bm{i}}$ is calculated for $\bm{i}$ a maximum distance from defects, here certain values of $r$ will correspond to the average including several defects simultaneously. In order to gain a measure of the average effect of defects on magnetic correlation we also simulated the system with randomly located defects at $\mathcal{T}=4.5$. For this case we observe that $\mathcal{G}_{\bm{i}}$ lies between the results for the ordered defects described above. 
\\
In order to understand this reduced magnetic order close to the defects we show the spatial dependence of  $\langle \eta_{\bm{i}} \rangle$ in Fig. \ref{eta_space} where we see that close to defects spins have a slightly increased average angle to the plane. In Fig. \ref{eta_space_rand} this dependence is shown with randomly located defects. We observe the same local increase in cone angle near to defects. The effect is especially  pronounced in the top left of the figure, where we observe a accumulation of defects associated with a region of significantly decreased canting. 
\\
When $w_d=8$ the average spacing between defects is large compared to the range of the local canting effect. In Fig. \ref{Corlen} we noted that, far from defects, short range correlations are comparable to those calculated for the isotropic system. In figures \ref{eta_spaceHD} and \ref{eta_space_randHD} we show the spacial dependence of $\langle \eta_{\bm{i}} \rangle$ with an increased defect density at $\mathcal{T}=4.5$. In the ordered case we have let $w_d=4$ and we see that the cone angle is no longer correlated with defect location. In contrast, when the same number of defects are randomly spaced as in Fig. \ref{eta_space_randHD}, clustering leaves areas where the cone angle remains small. 
\\
In Fig. \ref{CorlenHD} we plot $\mathcal{G}_{\bm{i}}(r)$ for the high density defects. Unlike the low density case  $\mathcal{G}_{\bm{i}}(r)$ does is not dependent on proximity to the ordered defects. We note also that the clustering effect means that the average short range correlation length is enhanced slightly when the defects are disordered. In all cases with high defect density the correlation length falls to zero at finite radius and so no in plane magnetization can form. In Table \ref{res} we give the cone angle and magnetization for the cases described here and note that within the precision of the simulation, despite the differences in morphology and correlation length, the strength of the magnetic ordering is dependent on the density rather than periodicity of the defects.
\begin{table}[htb]
\caption{Comparison of the properties of isotropic system, ordered defects, and random defects at $\mathcal{T}=4.5$.}
\begin{tabular}{lcc}
\hline\hline
\bf{System} & $ M_{\parallel} $ & $\eta$ \\
\hline
Isotropic\:&  0.463 & 0.529 \\
Low Density Defects\:& 0.363&0.556\\
Low Density Random Defects\:&  0.359 & 0.556 \\
High Density Defects\:& 0.064&0.622\\
High Density Random Defects\: & 0.080&0.614\\
\hline
\end{tabular}
\label{res}
\end{table}
\subsection{Fluctuations}
\begin{figure}[!htb]
  \centering
  \includegraphics[width=6cm]{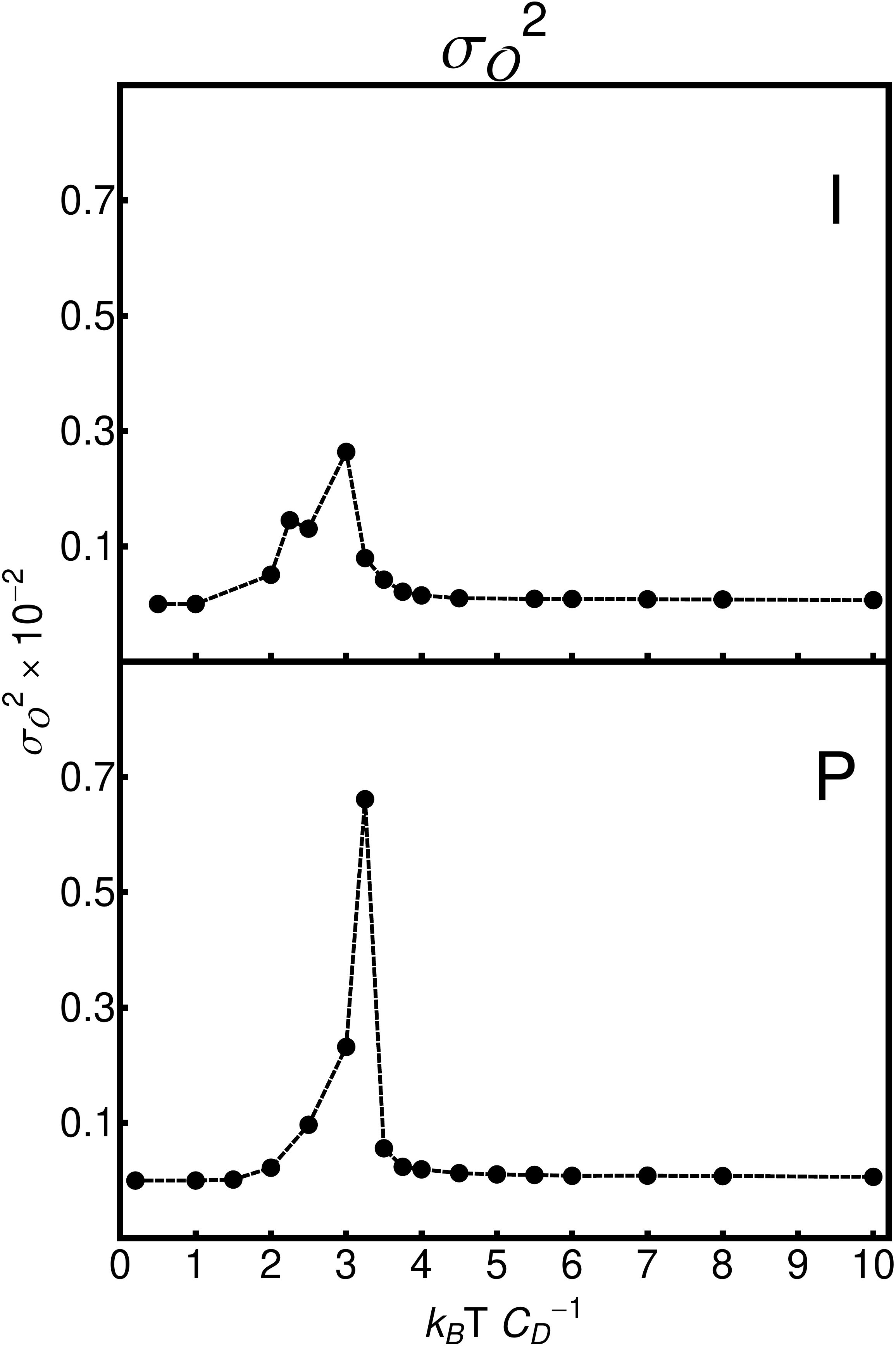}
  \caption[Variance of order parameter OT]%
  {$\sigma^2_{\mathcal{O}}$ as a function of $\mathcal{T}$ for the isotropic case (I) and the patterned case (P). }
\label{VOT}
\end{figure}

\begin{figure}[!htb]
  \centering
  \includegraphics[width=6cm]{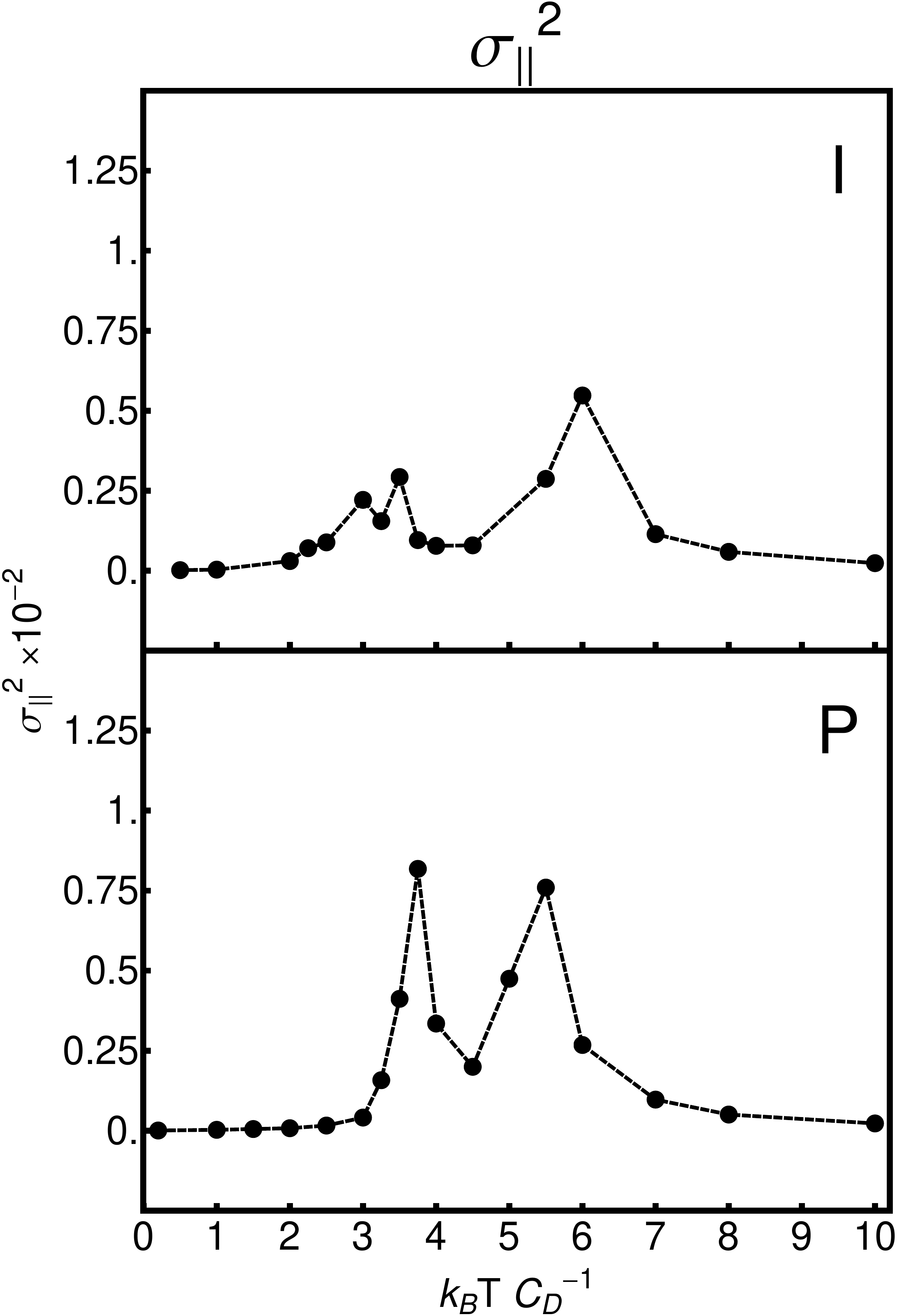}
  \caption[Variance of order parameter]%
  {$\sigma^2_{\parallel}$ as a function of $\mathcal{T}$ for the isotropic case (I) and the patterned case (P). }
\label{Vmag}
\end{figure}

\begin{figure}[!htb]
  \centering
  \includegraphics[width=6cm]{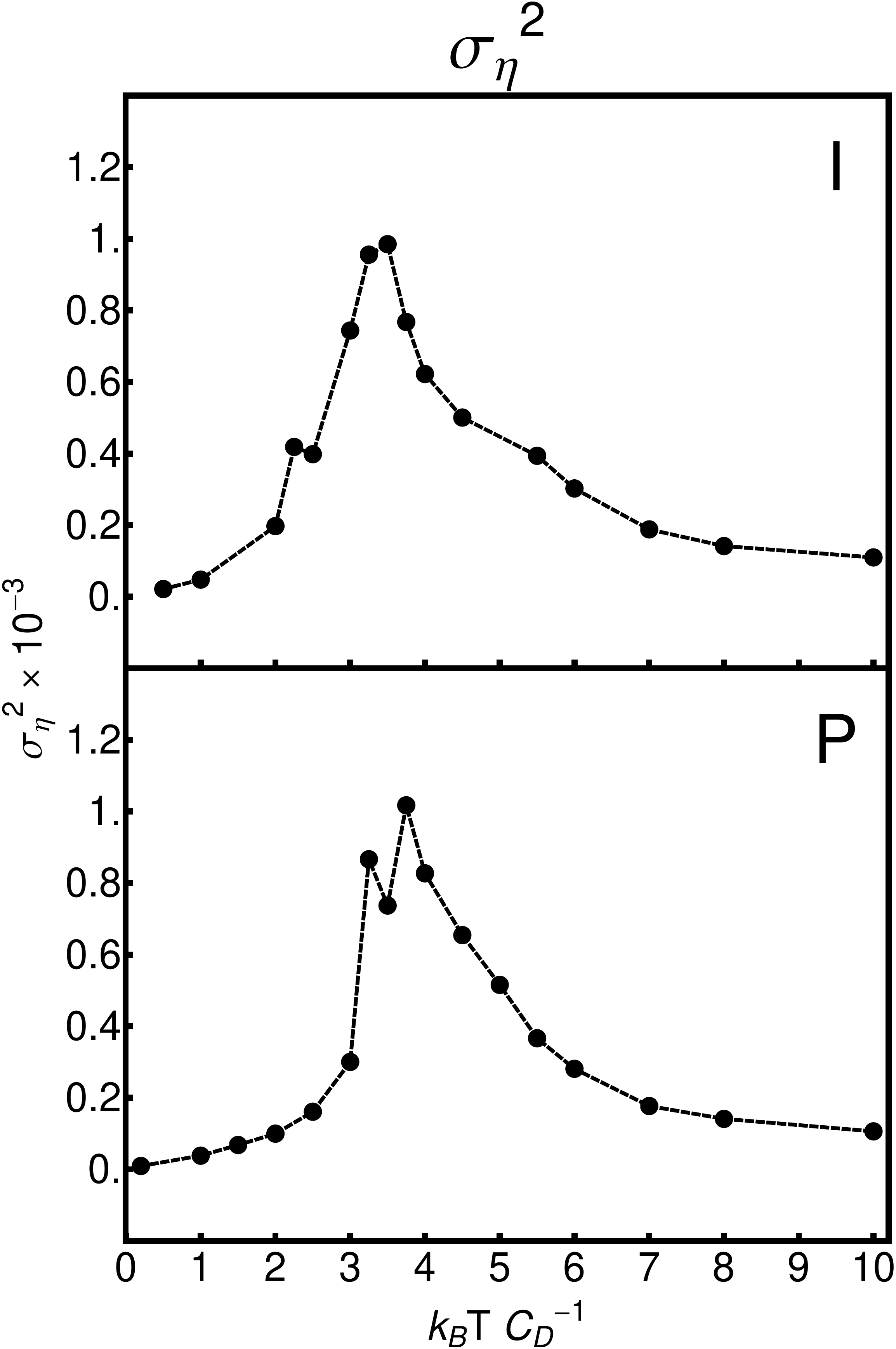}
  \caption[Variance of order parameter]%
 {$\sigma^2_\eta$ as a function of $\mathcal{T}$ for the isotropic case (I) and the patterned case (P).}
\label{Veta}
\end{figure}

We now consider the effects of ordered low density defects on fluctuations as a function of $\mathcal{T}$. We calculate the autocorrelation function
\begin{equation}
\sigma^2(X) = \langle(X- \langle X \rangle)^2 \rangle,
\end{equation}
 of the three order parameters $M_{\parallel}$, $\mathcal{O}^z$ and $\eta$, which we denote $\sigma^2_{\parallel}$, $\sigma^2_{\mathcal{O}}$ and $\sigma^2_\eta$ respectively.
In Fig. \ref{VOT} $\sigma^2_{\mathcal{O}}$is plotted as a function of $\mathcal{T}$, here we observe that the peak fluctuations occur at a higher temperature in the presence of defects corresponding to the stabilization of the striped structure.
\\
$\sigma^2_{\parallel}$ as a function of $\mathcal{T}$ is plotted in Fig. \ref{Vmag}. In both cases the fluctuations display two peaks corresponding to the creation and destruction of in-plane order. The low temperature peak is shifted towards higher $\mathcal{T}$ in the presence of defects.
\\
In Fig. \ref{Veta} $\sigma^2_\eta$ shows the same trend for both the patterned and isotropic cases, a broad peak with maximum occuring at $\mathcal{T}=4.5$ corresponding to the peak in-plane magnetic order.

\section{Conclusions and Comments}
Monte Carlo simulations have been used to investigate the effects of non-magnetic defects on the stripe melting and spin reorientation transitions. We have shown that the inclusion of non magnetic defects with spacing comparable to the natural stripe width affects the melting of stripes by creating pinning sites for domain boundaries favoring parallel alignment of stripes. At higher temperatures the two measures of the spin reorientation transition (reduction of the cone angle and the appearance of in plane magnetization) are reduced. In particular there is a spatial dependence of cone angle and correlation strength on proximity to a defect. Recalling that dipole coupling favors in plane alignments of spins \cite{PhysRevB.77.134417,PhysRevB.77.174415}, we surmise that the increased cone angle is due to the reduced dipole field near to the defects. The reduced dipole field increases the effective anisotropy near to the defects, suppressing canting away from perpendicular alignment.
The increased $\eta$ values reduce the size of the $\sin(\theta_{\bm{i}})\sin(\theta_{\bm{j}})\cos(\phi_{\bm{i}}-\phi_{\bm{j}})$ term in $\bm{s}_{\bm{i}}.\bm{s}_{\bm{j}}$, effectively reducing the exchange coupling of the $x$ and $y$ components of the spins, leading to the suppression of in-plane magnetization. 
\\
Here we have restricted our attention to point defects. Recently Van de Wiele et al. have performed a temperature independent micro-magnetic simulation of magnetization reversal in a sample with square holes \cite{wiele:053915}. Here the defects have dimension comparable to the spacing between defects. They find that the local shape anisotropy of the holes significantly affects the reversal mechanism. In light of these calculations it would be interesting in the future to consider to consider the melting problem on larger lattices where the effects
 of changing the size and shape of defects could be investigated.
\section*{Acknowledgments}
The benefited from fruitful discussion with Peter Metaxas. This work was funded by the Australian government department of Innovation, Industry, Science and Research, the Australian Research Council, the University of Western Australia and the Scottish Universities Physics Alliance.
\bibliographystyle{unsrt}	
\bibliography{NMD}
\end{document}